\documentclass[reprint,twocolumn,preprintnumbers,superscriptaddress,pre,amsmath,amssymb]{revtex4}
\usepackage{dcolumn}
\usepackage{hyperref}
\usepackage{color}
\usepackage{graphicx}
\usepackage{bm}
\usepackage[justification=raggedright]{caption}
\usepackage{amsmath}\begin{document}
\title{Anderson localization at the hybridisation gap in a plasmonic system}
\author{M. Balasubrahmaniyam}
\affiliation{Nano-optics and Mesoscopic Optics Laboratory, Tata Institute of Fundamental Research, 1, Homi Bhabha Road, Mumbai, 400 005, India}
\author{Ajay Nahata}
\affiliation{Department of Electrical and Computer Engineering, University of Utah, Salt Lake City, UT 84112, USA.}
\author{Sushil Mujumdar}
\email[]{mujumdar@tifr.res.in}
\homepage[]{http://www.tifr.res.in/\~mujumdar}
\affiliation{Nano-optics and Mesoscopic Optics Laboratory, Tata Institute of Fundamental Research, 1, Homi Bhabha Road, Mumbai, 400 005, India}

\date{\today}
\begin{abstract}

Disorder-induced Anderson localization in quasiparticle transport is a challenging problem to address, even more so in the presence of dissipation as the symptoms of disorder-induced localization are very closely simulated by the absorption in a system. Following up on recent experimental studies, we numerically study the occurrence of Anderson localization in plasmonic systems at terahertz frequencies. The low losses in the material at these frequencies allow us to separately quantify the localization length and the loss length in the system. We measure a non-monotonic variation of loss length as a function of disorder, and attribute it to the participation ratio of the localized modes and resulting light occupancy in the metal. Next, we identify a unique behavior of the gap state frequencies and the density of states under disorder. We observe that the maximally displaced gap state frequencies have a propensity to remain pinned to the frequency of the gap center. Even under strong disorder, the gap does not close, and density of states profile continues to remain peaked in the gap, unlike in conventionally studied disordered systems. The origins of this behavior are traced to the nature of the quasiparticle dispersion. In our case, the quasiparticles are identified to be hybrid plasmons generated due to the hybridization of surface plasmon polaritons at a metal-dielectric interface and cavity resonances at sub-wavelength apertures thereon. This situation is akin to the Kondo systems, where dispersive conduction electrons hybridise with a localized impurity state opening a hybridization gap. Our results provide new insights on the elusive problem of the interplay of loss and localization, and underlines interesting physics at the hybridisation gap in hybrid plasmonic systems.

\end{abstract}
\maketitle

\section{Introduction}

Anderson localization, proposed in the context of electronic transport\cite{anderson58}, is a wave interference phenomenon in disordered systems wherein transport is inhibited due to scattering. While the experimental achievement of localization has been done in various different scenarios such as sound waves\cite{weaver90}, matter waves\cite{billy08} and  light waves\cite{wiersma97}, the last system has seen an explosion of research activity in recent years\cite{segev13,wiersmareview,chabanov00,conti08,karbasi12,lahini10,lahini08,mookherjea08,sapienza10,schwartz07,topolancik07,szameit12, sperling12,nahata,liu14,randhir17}. Recent literature on Anderson localization has progressed well beyond the mere demonstration of localization, and has dealt with, among others, cavity quantum electrodynamics\cite{sapienza10}, quantum optics\cite{lahini10}, nonlinearity\cite{lahini08}, lasing\cite{liu14,randhir17} and temporal complexity\cite{randhir17} of Anderson localizing systems. In all these studies, the most commonly used structures consist of dielectric materials, presumably due to the absence of absorption that obfuscate the effects of localization\cite{scheffold99,weaver93}. Notably, despite such advances in investigations on localization, there are still no systematic studies on the physics of dissipative localizing systems. The primary detriment of dissipation is the fact that it induces an exponential decay in the transmitted field, which, otherwise, would be the signature of Anderson localization. Such exponential decays have been directly demonstrated in low-dimensional systems\cite{lahini08,schwartz07,nahata,randhir17}. An advantage of lower-dimensional systems is the relative ease of obtaining localization, as against three dimensional structures where a critical degree of disorder is necessary. A particularly interesting route for achieving low-dimensionality is surface plasmons, which are inherently low-dimensional entities. Indeed, several works have addressed transport in disordered plasmonic structures such as random planar metal dielectric composite\cite{gressilon99,stockman01,markel06}, metal nanosphere arrays\cite{markel07},  fractal and random nonfractal clusters\cite{stockman97,krachmalnicoff10,caze13}, subwavelength hole arrays on metal films\cite{randhir_optcomm,nahata} etc. However, conventional plasmonic materials such as silver and gold, at conventional plasmonic frequencies, have a major issue related to the very strong absorption.

Towards a meaningful investigation of dissipative localizing systems, a thoughtful choice of materials needs to be made such that the inherent dissipation is not forbidding. To that end, terahertz frequencies in metallic systems are promising candidates. The terahertz range of frequencies lies at the boundary between low-frequency and high-frequency radiation. The latter radiation realizes highly dissipative plasmons on metal surfaces, while the former (low-frequency) does not penetrate the metal due to its diverging conductivity.  At terahertz frequencies, real metals respond with large but finite conductivities\cite{pandey13}, and hence can sustain surface-bound plasmons with a low loss. In this case, the presence of a subwavelength structure realizes hybrid plasmons\cite{pendry4}. We have recently experimentally demonstrated Anderson localisation in a hybrid-plasmonic system at terahertz frequencies\cite{nahata}, and directly imaged the localized wavefunctions. The structure comprised waveguides formed by coupled subwavelength holes in a thin metal sheet. We measured the dissipation length in the structures, and showed that the eigenfunction decay primarily arose through localization. While this was the first such demonstration in the terahertz frequency domain, the experimental limitations restricted certain measurements despite state-of-the-art techniques. In the finite-sized system, the transmission bands comprise spectrally-separated collective modes, which undergo localization under disorder. These individual modes within the bands could not be frequency-resolved and only the modes outside the Bragg frequency could be measured where no bands exist. The rarity of such modes preclude the identification of any systemic trends. Furthermore, the dissipative behavior associated with the \emph{individual} localized modes cannot be identified. This experimental study is the motivation of the current theoretical work. In this study, we approach the plasmonic localization from the physics of dispersion of hybrid plasmons. To that end, we use a finite-element eigensolver which computes the eigenvalues and eigenfunctions of the structure. We first identify the transmission bands of the infinite periodic device and then resolve one band into the fundamental resonances of a finite structure as used in experiments. Next, we analyze the effect of disorder on these modes at a frequency resolution inaccessible in the experiments and diagnose the frequency shifts, localized eigenfunctions and the losses separately.  Finally, we observe that the unique dispersion in the system manifests unusual behavior of the density of states at the bandedges and in the bandgap. This is elucidated through a comparative study of the disorder-induced localization of hybrid plasmons with conventional Anderson localization in coupled dielectric cavities. This paper is arranged as follows: In section II, we describe the physical structure and the generation and dispersion of the hybrid plasmons. Section III is aimed at explaining the experimental results, but at spectral resolutions inaccessible in experiments, and thereby assess the loss-localization behavior at an individual modal level. Section IV takes the analysis into the unusual behavior of the bandgap, and its origins in the dispersion of plasmons. Section V closes with conclusions.

\section{Physical structure}

Figure 1(a) shows the plasmonic system under investigation. It consists of a 1-D periodic rectangular through-hole array etched in a $500~\mu$m thick stainless steel sheet. The inset in Figure 1(b) shows the unit cell of the array. The dimensions of the hole $h$, $s$ and $a$ determine the cavity resonances. They cannot couple with each other directly because of the metal region between the cavities. The coupling is accomplished via a propagating linear-dispersive plasmon parallel to the metal surface. Finite element analysis of the fields in the structure were carried out using the software COMSOL Multiphysics, using a measured dielectric behavior of stainless steel at terahertz frequencies\cite{pandey13}.

The technique involves computing the eigenmodes of any particular structure defined on a virtual grid. This is achieved by solving the following eigenvalue problem derived from the Maxwell's equations
\begin{eqnarray}
\{\nabla \times \mu^{-1} \nabla \times\} \bf{E}= \epsilon ~ \omega^2/c^2\textbf{E}
\label{eq:one}.
\end{eqnarray}
where $\mu$ and $\epsilon$ are the relative permeability and permittivity defined at each point of the grid, which realizes the structure. The eigenvalue equation is solved employing appropriate boundary conditions. Thus, the eigenfrequencies $\omega + i\Gamma$ and the corresponding eigenfunctions of the system are obtained. The imaginary part of the eigenfrequency ($\Gamma$) yields the temporal decay which is converted to loss length as $c/\Gamma$. Both the periodic and the disordered systems will have their characteristic loss lengths, as determined by the field in the metal. In case of an infinite periodic system, a single unit cell is discretized and a periodic boundary condition along one direction is applied, along with a perfectly-matched layer (PML) along the other two directions. In case of the finite ordered/disordered systems discussed later, the entire structure is discretized and enclosed within PML’s. The virtual grid in this computation is implemented by adaptive meshing, whose maximum and minimum lengthscales were kept equal to, respectively, one eighth and one tenth of the lowest wavelength considered in the THz spectra.

\begin{figure}[h]
\includegraphics[width=8.5cm]{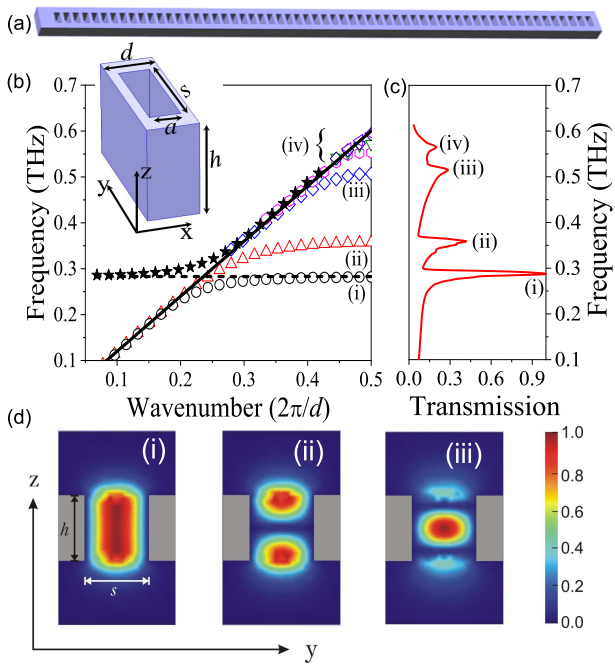}
\caption{\label{fig:epsart1}(a) Schematic of the 1-D rectangular hole array in the periodic configuration. (b) The inset shows the schematic of the unit cell, with $a= 150~\mu$m, $h= 500~\mu$m, $s=500~\mu$m, and $d= 250~\mu$m. The main plot shows the calculated bandstructure of the infinite periodic rectangular hole array. The dashed line represents the first uncoupled-cavity resonance and the tilted solid line is the plasmon dispersion. Their hybridization results into a lower (marked as (i)) and upper band, shown by the $\circ$ markers and $\star$ markers respectively. The $\bigtriangleup$ markers and $\diamond$ markers show the lower bands (marked as (ii) and (iii)) corresponding to the hybridization with the second and third order cavity resonances. See text for (iv). (c) Normalized transmission spectrum of the periodic system as computed from the bandstructure in (b). The peaks corresponding to each band are marked. (d) Simulated cross sectional intensity distributions in the yz plane at the center of a subwavelength hole, for the first (i), second (ii) and third (iii) bands. Grey rectangles depict the metal region.}
\end{figure}

Under periodic boundary conditions, the bandstructure calculated from the eigensolutions of the infinite periodic array is shown in Figure 1(b). The dispersion of the resulting band is essentially determined by the hybridisation of the spatially localized cavity resonance, represented by the horizontal dotted line and the surface plasmon, whose dispersion is shown as a solid straight line. We note that, at these frequencies, the surface plasmon dispersion almost overlaps with the light line.  The anti-crossing resulting from the hybridisation realizes a bandgap. An example of the anti-crossing and gap formation is shown in Figure 1(b), where the $\circ$ markers and $\star$ markers designate the lower band (marked as (i)) and upper band realized due to the hybridisation with the 1$^{st}$ cavity resonance. The $\bigtriangleup$ markers and $\diamond$ markers show the lower bands (marked as (ii) and (iii)) corresponding to the hybridization with the second and third order cavity resonances. The upper bands form above the light line and are radiative, while the lower bands are non-radiative and hence support surface transport. The frequencies at which band gap occurs,i.e. the anti-crossing points, are completely determined by the resonant frequencies of the cavity and are independent of the periodicity. The periodicity $d$ determines the Bragg frequency of the system, given by $c/2d$, which, in our case, is 0.6~THz. The bands exhibit a linear dispersion close to $k = 0$ and saturate close to the edge of the Brillouin zone,i.e., $k=\pi/a$. Close to the Bragg frequency, an indistinct cluster of bands (labeled as (iv)) is formed through the coupling of various higher order resonances of the cavity. Using the real and imaginary frequencies extracted from the eigensolutions, the longitudinally transmitted intensity spectrum is calculated as the sum of Lorentzian peaks located at the respective frequencies of the system. As shown in Fig. 1(c), the spectrum shows asymmetric peaks corresponding to each individual band. A broad peak, corresponding to the collection of bands, occurs close to the Bragg frequency. The peak amplitudes fall with increasing  order of the band, evidently due to the fewer modes in the higher bands. Fig. 1(d) shows their intensity distribution in the unit cell, computed at the yz plane at the center of a hole. Penetration of the terahertz intensity in the higher order bands (iv, not shown here) in air is extremely weak, and hence, these bands escape experimental detection methods that rely on index modulation of electro-optic crystals\cite{nahata}.

We note that this system has qualitative equivalence in Kondo systems, wherein magnetic impurities in a metallic crystal create a hybridisation of the conduction band of the $d$ orbital, which is dispersive, and the spatially localized $f$ electrons, which has a zero-dispersion profile. This hybridisation opens up a gap. In our system, the dispersion-less band is simply provided by the cavity resonance. The dispersive band is provided by the surface plasmon polariton (SPP) dispersion, which, at terahertz frequencies, is extremely close to linear dispersion that almost overlaps the light line. In this paper, we call the hybridised band as the hybrid plasmon. Interestingly, for higher conductivities such as in perfect-metals or at microwave frequencies, the linear dispersion represents the low-frequency limit of a SPP wave, known as the Sommerfeld-Zenneck wave. This wave essentially constitutes an electric field oscillation in the dielectric medium propagating parallel to metal-dielectric interface. In this case, precisely because the coupled dispersion behavior is similar to surface plasmons, they have been identified as spoof plasmons\cite{pendry4, garcia5}. This unique transversally-confined yet laterally-propagating plasmon has attracted a lot of attention in recent years, and has extended plasmonics into the low frequency regime\cite{hibbins05,song09,song11,erementchouk16,joy17}. The analysis that we present in this paper is expected to be valid also for the spoof plasmons.

\section{Dissipation and localization}

\begin{figure}[h]
\includegraphics[width=8.5cm]{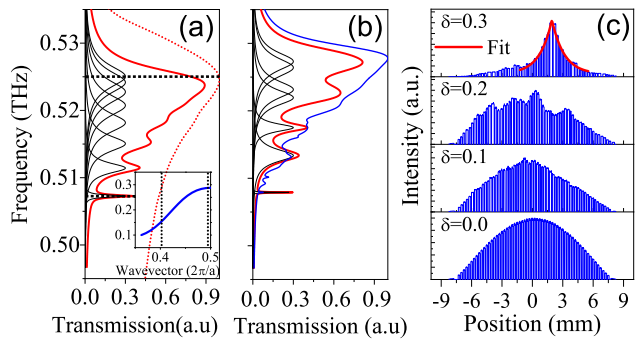}
\caption{\label{fig:epsart2}(a) Transmission peak of band (iii) of the finite periodic system (solid red line) and infinite periodic system (dotted red line). Black spectra indicate the underlying resonances of the system. Inset shows the fraction of the field inside the metal. (b) Effect of disorder: Black spectra show the resonances of the disordered array. The red spectrum shows the transmission peak composing the resonances. Blue spectrum shows the peak averaged over 40 configurations. (c) Intensity (on a linear scale) as a function of position. The plot illustrates the evolution of the fundamental resonance with increasing disorder, showing the exponential decay in the wings due to localization.}
\end{figure}

In a finite hole array as is used in any experiment, each band will split into multiple collective modes determined by the length of the system. To recreate the experimental situation, we investigated a finite-sized hole array and analyzed the resulting discrete collective modes. In this case, the simulated structure consisted of 60 holes, creating a length of 15 mm, further enclosed by metallic padding of 2.5 mm on either side. This realized the total structure of length 20~mm (system size $L$). The transverse dimensions remain the same as the infinite structure. We discuss here the behavior of band (iii) as it provides clutter-free individual modes, however, the behavior is the same for all bands. Figure 2 discusses the spectral behavior at a high frequency resolution typically inaccessible in the experiments. In Figure 2(a), the black spectra show discrete Lorentzians which represent the individual collective modes of the finite system that add together to form the continuous spectrum (solid red line). For comparison, the dotted red line shows the spectrum for the infinite array. Interestingly, the mode at the bandedge (marked by the horizontal dotted line) is essentially the first-order mode of the finite system and the order increases further away from the bandedge. Correspondingly, the width of the Lorentzian reduces towards the higher order. This is in contrast to dielectric (loss-less) finite periodic systems, where the first order mode shows the highest quality factor and hence the lowest width. This originates from the metallic losses associated with the mode. We analyzed the light intensity resident in the metallic region for every mode, and the result is shown in the inset of Fig 2(a). Clearly, the intensity inside the metal is maximum at the bandedge and reduces monotonically further from it. The range of wavevectors shown here covers all the discrete modes shown in the main plot, and the behavior was observed to be monotonically decreasing for further $k$ (not shown). Thus, the metallic losses dominate for the modes away from the light line, whereas the mode along the lightline incur lesser metallic losses, resulting in the broadening of modes closest to the bandedge.

Next, disorder was introduced in the finite system by the following procedure. The position of each resonator is displaced by an amount determined by a uniformly distributed random number in the set $[-\delta*d/2,\delta*d/2]$, where $\delta$ ranges from $0\leq\delta\leq1$.
For each disorder strength $\delta$, 40 configurations were simulated. Figure 2(b) discusses the behavior for an individual configuration at $\delta = 0.3$. In general, the individual discrete modes were observed to blueshift towards and across the bandedge. The amount of the shift is variable for different modes. For instance, the bandedge mode at 0.525 THz shifted to 0.531 THz incurring a shift of 0.006 THz, compared to the highest order mode that shifted by $\sim0.002$~THz. The spectral width of the modes also changes upon introduction of disorder, with some modes broadening and others narrowing down, depending on the metallic losses in the modes. The total spectral line (shown in red) is effectively blue-shifted, but the shift is of a small magnitude in comparison to its width, and hence can be expected to escape experimental detection. The blue line shows the configurationally-averaged spectrum over 40 configurations.

Anderson localization of the modes can be seen in the spatial distribution of intensity, which shows a clear transformation. Figure 2(c) shows the intensity distribution for the fundamental mode (bandedge mode) for four disorder strengths. For the periodic structure, the distribution shows a perfect sinusoidal with a single maximum as expected in the first order mode of a finite system. However, as disorder increases ($\delta = 0 - 0.3$), the distribution transforms into one with exponential tails in either wings of the form $\exp -(|x - x_0|)/\xi$ where $\xi$ is the decay length. Such a spatial exponential decay is a clear indication of Anderson localization. On fitting the exponential for the bandedge mode, the $\xi$ turns out to be 4.8 mm. This computed spatial distribution has no contribution from the intrinsic (metallic and radiative) losses of the system and the calculated decay here is purely due to the localization. The computed imaginary part of the eigenvalue of the mode yields the decay length of 13.7~cm, which is a factor of $\sim25$ larger than the localization length $\xi$.

\begin{figure}[h]
	\includegraphics[width=8.5cm]{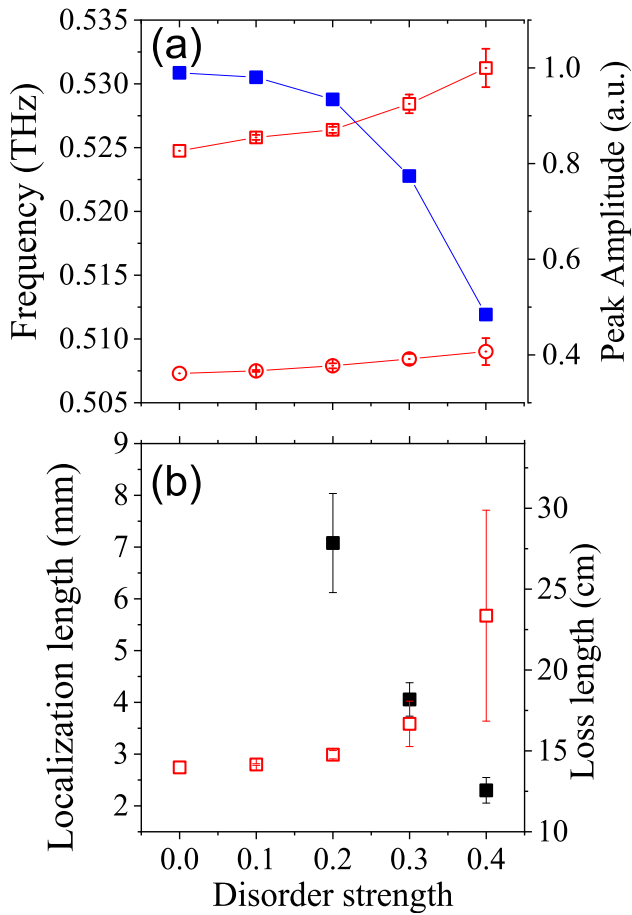}
	\caption{\label{fig:epsart3}(a)Red $\Box$'s show the translation of the resonant frequency (left Y axis) of the fundamental resonance, while the red $\circ$'s show the same for the eighth order resonance. The resultant peak in the spectrum shows a decreasing peak amplitude (blue $\blacksquare$ markers, right Y-axis). (b) Average localization length (black $\blacksquare$ markers) reduces with increasing disorder. The average metallic loss lengths  (red $\Box$ markers, right Y-axis) increase, and the fluctuations in the losses also increase.}
\end{figure}

These numbers, however, only represent a single configuration, and the complete picture is only seen after configurational averaging. Figure 3(a) quantifies the shift in the resonant frequencies over 40 configurations at $\delta = 0.3$, where the $\Box$ markers depict the fundamental mode (bandedge mode) and the $\circ$ marker depicts the eighth-order mode. The dissimilar shifts result in widening of the spectral peak. Further, the amplitude of the peak also drops correspondingly, as seen in the plot with blue $\blacksquare$ markers. Figure 3(b) shows the behavior of the localization length and the loss length of the bandedge (first-order) modes over 40 configurations.  In this work, we focus only on the bandedge mode because the loss is maximum for the bandedge mode, and it reduces further from the bandedge. So the bandedge mode is sufficiently representative of the systemic loss. Clear exponential fits were only achievable at and above $\delta = 0.2$, shown in black $\blacksquare$ markers. The average localization length drops with increasing disorder, as is common with any Anderson-localizing system. The fluctuations of the localization length (quantified by the error bars signifying the standard deviation in $\xi$) also reduce with disorder strength. The red $\Box$ markers indicate the loss length of the bandedge modes, and is seen to increase with disorder indicating reduced losses. The fluctuations in the loss, however, increase with disorder. The reason thereof is that, as in dielectric systems, the quality of localized modes depends upon their location vis-a-vis the edges of the system. Some modes manifest close to the edge and hence are strongly coupled to the environment, and incur larger losses, compared to those that are centrally located. This location-dependent loss is over and above the metallic losses that depend on the field extent inside the metal, which, in turn, is determined by the strength of localization of the mode. This study emphasizes the interesting interplay of loss and localization in hybrid plasmonic systems.

\begin{figure}[h]
	\includegraphics[width=8.5cm]{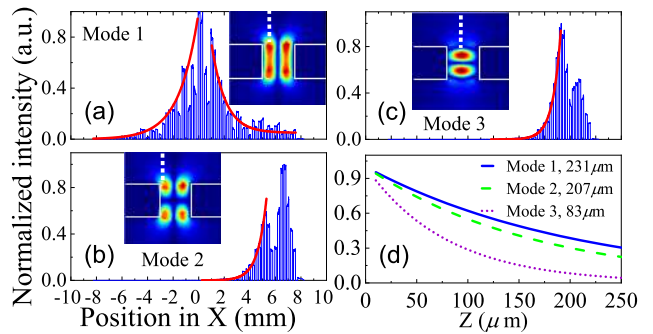}
	\caption{\label{fig:epsart4}Intensity distribution of three localized modes extracted beyond the Bragg frequency under disorder of $\delta = 0.3$. The frequencies of the localized modes 1, 2, and 3 are $0.6121$~THz, $0.6124$~THz, and $0.6125$~THz respectively and the losses are $18.7$~ns$^{-1}$, $19.5$~ns$^{-1}$, and $19.6$~ns$^{-1}$ respectively. The corresponding $\xi$ are 1.8~mm, 1.2~mm , and 0.9~mm. Each inset shows the corresponding cross-sectional field distribution, in the same perspective as in Fig~1(d). The field decay into air (along the white dotted line) is shown in (d), and yields a decay length of 231~$\mu$m, 207~$\mu$m, and 83~$\mu$m, respectively, for the three modes.}
\end{figure}

At this stage, we briefly address the experimental measurements of Anderson-localized modes in such a system. As mentioned earlier, the experimental techniques in the terahertz domain employ electro-optic devices that inherently determine the capability of detection of the modes. In the transmission bands below the Bragg frequency, the modes are all spectrally and spatially bunched too close for detection. In this region, broad features such as spectral shifts and broadening, as observed above, can be noticed in experiments. The situation, however, can be different in the region of the Bragg frequency, where there are no pre-existing modes. At an appropriate disorder strength, some modes from the bands transit beyond the Bragg frequency. Figure 4 depicts three localized modes, each one being the bandedge mode of the respective band, in a single configuration of $\delta =  0.3$, all observed outside the Bragg frequency. Each mode is the highest-quality factor mode in its respective band, whose field distribution is shown (inset) in the same perspective as in Fig~1(d). Subplot (d) shows the decay of the field into the air, along the white dotted lines shown in each inset. Clearly, the modes 1 and 2 have a better penetration into the air above, compared to the mode 3, which is more tightly bound to the surface. However, mode 1 has the minimum loss among all four, primarily  due to its central placement compared to the other modes that are closer to the edge and hence leaky. Under such conditions, mode 1 will be preferentially detected, providing a clean measurement of an isolated Anderson localized mode beyond the Bragg frequency. Such a clean isolated mode was indeed observed in the experiments outside the Bragg cutoff\cite{nahata}. The localization length for this mode is calculated to be $1.8$~mm. Thus, the ratio of $\xi/L$ is 0.09 at a disorder strength of $0.3$, which is very much in agreement with the experimentally measured $\xi/L$, which was 0.1 at $\sigma = 0.25$. It must be emphasized that the localized modes will also exist in regions below the Bragg frequency, but a clean and unambiguous detection is possible only above the Bragg frequency.

\section{Localized modes in the hybridization gap}

While weak disorder mostly realizes modes in the vicinity of the bandedge, stronger disorder tends to create modes deeper into the bandgap. So, we extend our study to disorder strengths ranging from $\delta = 0.3 $ to $ 0.9$, whose observations are provided in Figure 5. We first quantify the loss at strong disorder, and study its correspondence with the modal extent. We note that the localization length $\xi$ only characterizes the decay of the tail, and not the total spatial extent of a mode. Therefore, we measured a mode size, equivalent to the inverse of the inverse participation ratio (IPR). The IPR is defined as $ L\times\frac  {(\sum ^N_{i=1} \psi ^2_i)^2} {\sum ^N_{i=1} \psi ^4_i} $, where $\psi_i$ is the value of the wave function at the $i^\text{th}$ site and $L$ is the system size. Figure 5(a) shows variation of the mode size (black $\Box$'s) and the average modal losses (red $\blacksquare$'s) that reduces with increase in disorder strength, averaged over 40 configurations. Not only does the averaged loss, but also the fluctuations in the loss show a strong correlation with the mode size. At low disorder $\delta =< 0.2 $, the mode size varies only weakly with disorder. At these disorder strengths, the modes typically occupy the entire sample size, and hence the mode size is determined by the system size, which becomes the limiting length-scale here. At intermediate disorders ($\delta = 0.3 $ to $ 0.7$) the mode size drops rapidly as the localization length reduces. Since the modes that occur at these disorder strengths are tighter, their peak positions largely vary from one configuration to another. Whenever the modes are closer to the edge, the mode size is smaller compared to those peaking at the center. As a consequence, the mode size and modal loss show stronger fluctuations as seen from Figure 5(a) close to  $\delta \sim 0.3 $. At higher disorder, $\delta > 0.4$, the mode size tends to saturate, driven by the limiting lengthscale of the average inter-hole separation. Interestingly, at high disorder strengths,  the localization lengths are approximately an order of magnitude smaller than the system size, and a majority of the modes occur in the bulk of the system away from the boundary. As a result, the fluctuations in the mode-size and loss also reduce.

\begin{figure}[b]
	\includegraphics[width=8.5cm]{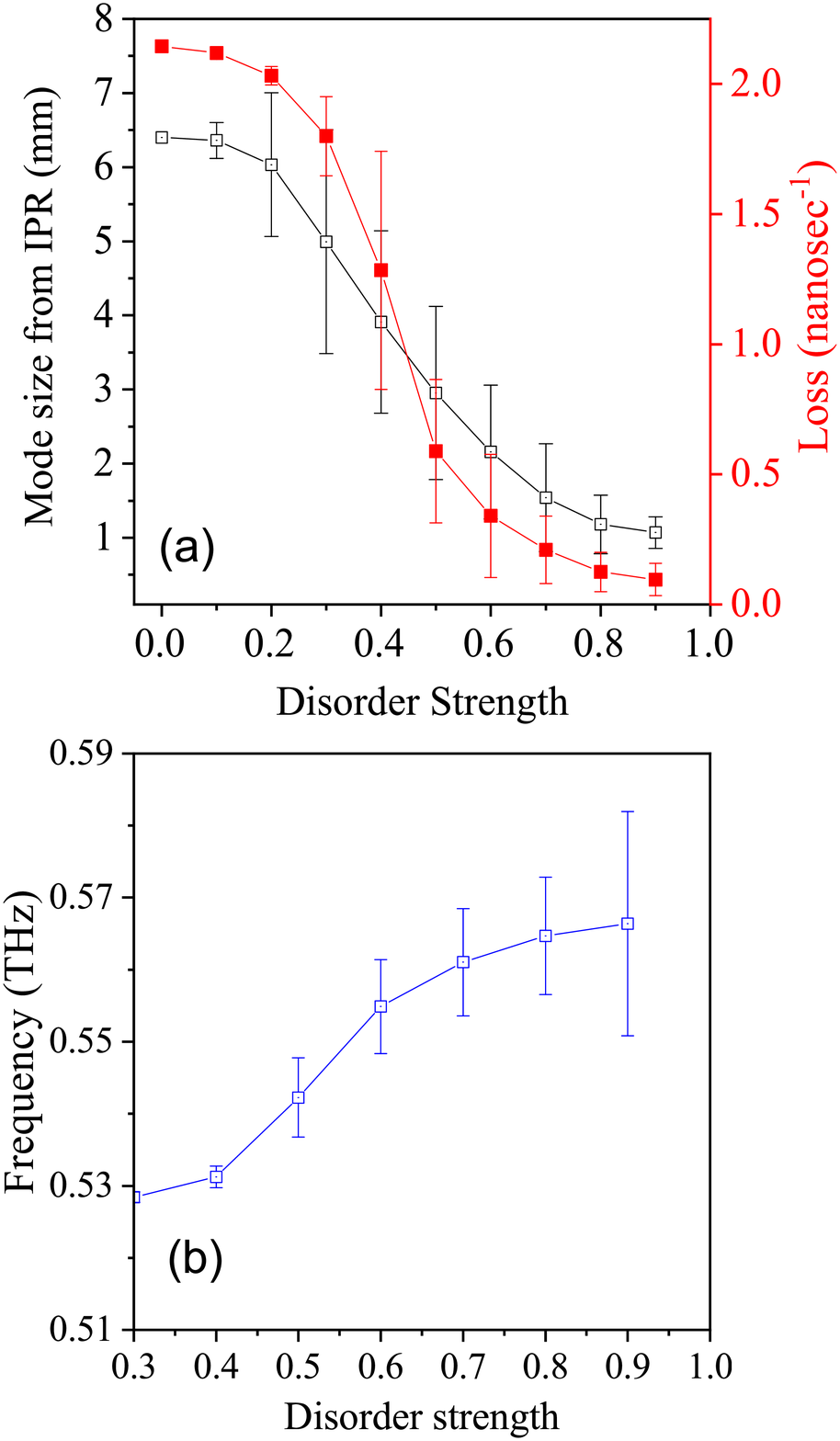}
	\caption{\label{fig:epsart5}(a) Black $\Box$'s show the mode size that reduces with increasing disorder. Red $\blacksquare$'s indicate the average modal losses whose variation follows the mode size. Error bars show the standard deviation in the parameters, that were configurationally averaged over 40 configurations. (b) Blue $\Box$'s show the migration of the deepest gap mode as a function of the disorder strength.}
\end{figure}

Figure 5(b) discusses the frequency of the gap mode that is generated deepest into the bandgap, beyond $\delta = 0.3$. The data are averaged over 40 configurations. At intermediate disorder, migration of the frequency into the bandgap accelerates. However, at higher disorder, the shift slows down and the resonant frequency of the deepest mode actually saturates. In this case, the average frequency at which the mode gets pinned is observed to be $\sim0.57$ THz. Such pinning of modes at high disorder strength is not observed in usual PARS systems made of coupled dielectric cavities or waveguides, where the band originates from direct coupling (either evanescent or radiative) of cavity resonances. To understand this behavior, we compare a hybrid plasmon band with that of a coupled dielectric cavity array under disorder.

In order to compute the density of states, a larger system size is necessary to be simulated. Hence, we performed these calculations using the tight-binding model \cite{xie12}. In this calculation, the actual shape or size of the resonance is not relevant, but just the resonant frequency is invoked in a Hamiltonian formalism. The Hamiltonian consists of cavities coupled with each other through a collective band (representing the plasmon) to simulate a single band hybrid plasmonic cavity excitation as
\begin{eqnarray}
H_{hp}&=&\sum_{i} t_{i,i+1} c^{\dagger}_{P_{eff}} (x_i) c_{P_{eff}} (x_{i+1}) \nonumber\\
 &  & + \sum_{i} E_{cav} (x_i) f^{\dagger}_{cav} (x_i) f_{cav} (x_i) \nonumber\\
&  & + \sum_{i} V_i c^{\dagger}_{P_{eff}} (x_i)  f_{cav} (x_i)+ H.c.
\label{eq:two}.
\end{eqnarray}
where $c_{P_{eff}}  (x_i )$  is the annihilation operator of the effective Hamiltonian for the plasmon, $f_{cav}$ is the annihilation operator associated with the cavity resonance, ‘$t$’ is the hopping strength and $E_{cav}$ is the cavity resonance frequency. $V_i$ represents the hybridization amplitude of the plasmon with the cavity at each site. This Hamiltonian is matricized into a finite near-diagonal matrix consisting of 1000 unit cells. Disorder in the spacing between the cavities is invoked as random hopping probabilities or the off-diagonal terms in the above Hamiltonian. The randomness in the hopping parameter was calculated as follows. First, in the finite system discussed in Section III, we calculated the maximum and minimum overlap integral of the field in any configuration, signifying the nearest and the farthest cavities respectively. A uniformly distributed random number between these two values was employed as the hopping parameter at each site. Thus, a homogeneity was maintained between the definition of disorder strength in the earlier computation and the current tight-binding one. The $E_{cav}$ is kept constant in this simulation. One thousand configurations were implemented at each disorder strength for configurational averaging. The matrices with different disorder strengths are diagonalized to obtain the eigenfrequencies and eigenfunctions. The density of states (DoS) are calculated from the eigenvalues.
\begin{figure}[h]
	\includegraphics[width=8.5cm]{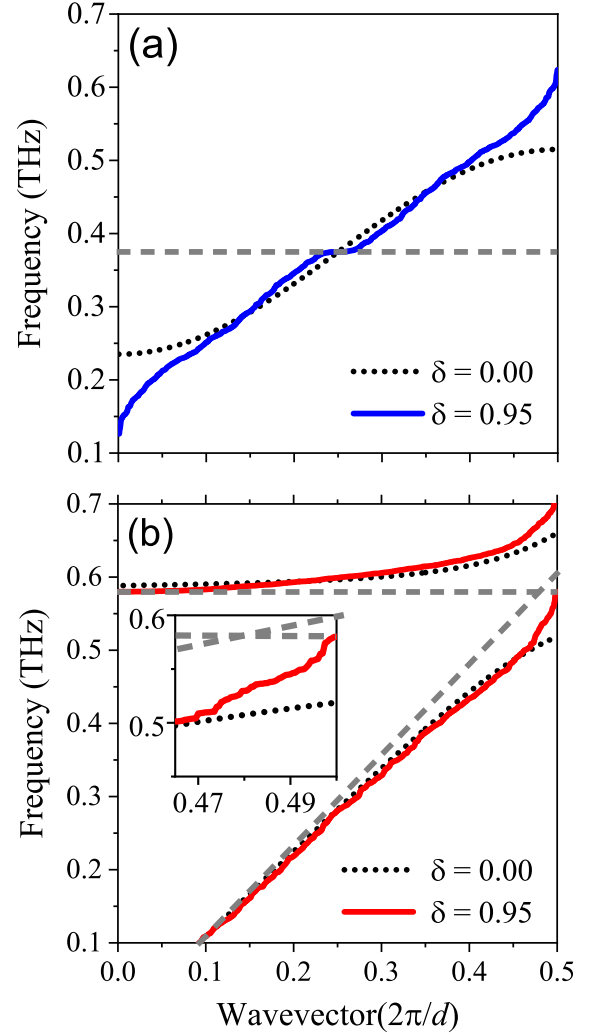}
	\caption{\label{fig:epsart6} (a) Dielectric coupled resonator system: Horizontal dashed line represents the cavity resonance of a single dielectric resonator. Dotted black line depicts the band arising due to the coupling of multiple such resonators, as computed from the tight-binding model. Blue continuous line shows the effect of 95\% disorder on the band. (b) Plasmonic coupled resonator system: Horizontal dashed line shows the individual cavity resonance, while the angled dashed line indicates the surface plasmon dispersion. Dotted black lines show the hybridized bands creating a bandgap at the resonance frequency. Continuous red lines indicate effect of disorder on the bands, where the lower band is limited by the resonance frequency, as emphasized in the inset.}
\end{figure}
For comparison, a dielectric cavity array was simulated by the standard tight-binding hamiltonian as
\begin{eqnarray}
H_{dc}&=&\sum_{i}  p_{i,i+1}  c^{\dagger}_i  c_{i+1} + H.c.
\label{eq:three}.
\end{eqnarray}
where the $c_i$ is the annihilation operator of the coupled cavity mode and ‘$p$’ is the hopping probability between the cavities. The comparison is provided in Fig. 6. In a dielectric cavity array (Fig~6(a)), the bandgap and the bandedges are spectrally distant from the resonant frequency of the cavity, which is positioned deep inside the band. Here, the dashed line represents the cavity resonance whereas the dotted black line shows the band obtained by solving the tight-binding model for the periodic system.  Upon introduction of disorder, the bandedge modes migrate into the gap (blue continuous line in Fig. 6(a)). The behavior of the hybrid plasmon is fundamentally different from this. The hybridisation of the cavity resonance and the SPP dispersion line realizes the avoided crossing at the resonant frequency of the cavity. Hence, the bandgap or the hybridisation gap is formed exactly at this resonant frequency as seen in Fig.6(b). The bands will reach  the resonant frequency of the cavity resonance asymptotically far away from the $k$ values of that of the anti-crossing point, and will close the gap. However, the finite $k$ space of the periodic system precludes the closing of the gap and results in the hybridisation gap at the cavity frequency. Upon introduction of disorder, the deepest gap modes do migrate into this gap (red continuous lines in Fig. 6(b)). However, even at highest disorder strength ($\delta = 0.95$), the migrating modes only reach the cavity resonant frequency. The inset emphasizes the Brillouin zone boundary. Figure 7 compares the deepest gapmodes in the dielectric and hybrid plasmonic systems. In the latter system (red circles), the deepest modes are pinned below the resonant frequency of the cavity resonance. The resonant frequency of the cavity here is $0.58$ THz, and the frequency of the gap mode is seen to get pinned at $~ 0.57$ THz. In comparison, the migrated modes of the dielectric system (blue squares) do not get pinned down.
\begin{figure}[h]
	\includegraphics[width=8.5cm]{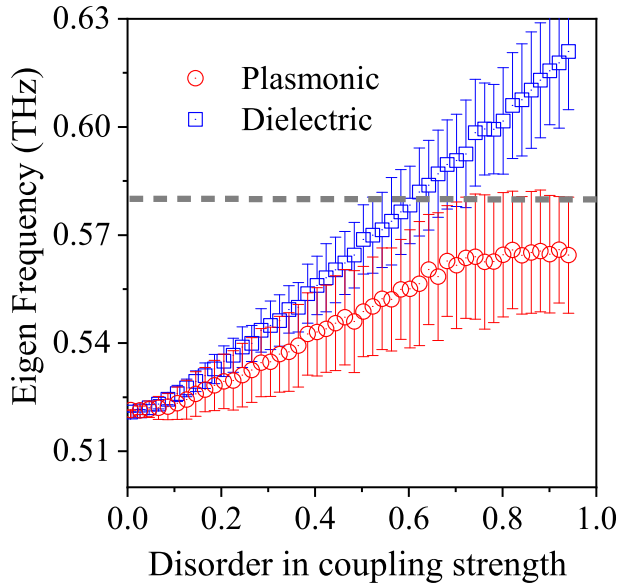}
	\caption{\label{fig:epsart7} Blue squares show the frequency of the mode localized deepest in the bandgap for the dielectric system for increasing disorder. Red circles show the same for the plasmon system, where the saturation of the mode below the frequency of the cavity resonance  (gray dashed line) is clearly seen.}
\end{figure}

\begin{figure}[h]
	\includegraphics[width=8.5cm]{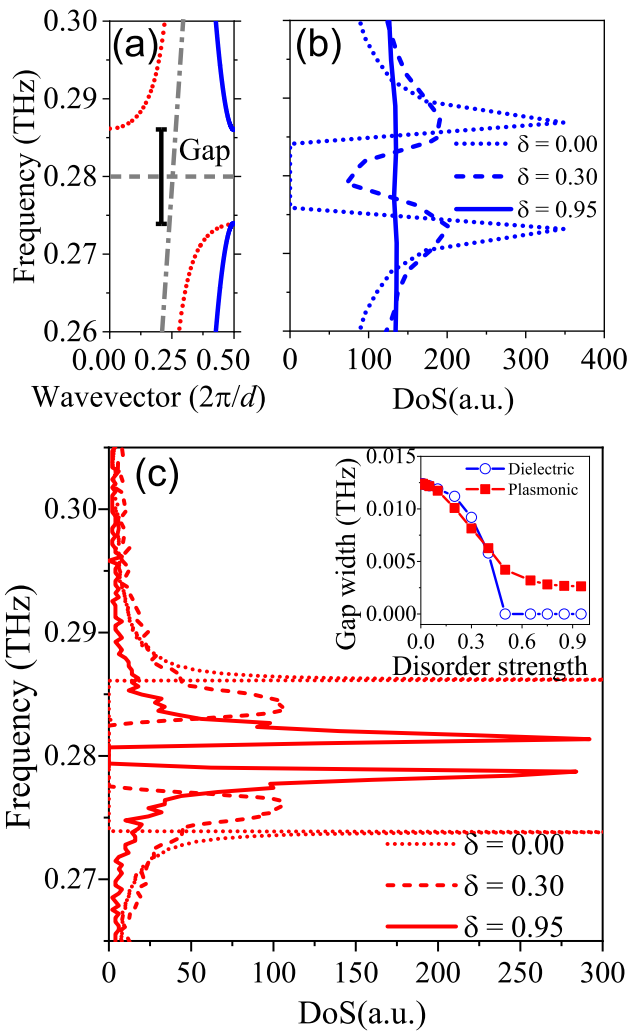}
	\caption{\label{fig:epsart8} (a) Dotted red lines show the hybrid plasmonic bands corresponding to the $1^{st}$ plasmon band in Fig~1. Solid blue lines indicate dielectric bands for a system designed for a matching bandgap. (b) Evolution of the density of states (DoS) of the dielectric system with disorder, showing a smoothening of the Van Hove singulariy and flattening of the DoS profile. (c) DoS for the  hybrid plasmonic system shows that the two peaks at the bandedges never merge under disorder and the DoS at the gap center (matching the cavity resonance frequency 0.28 THz)remains zero. Inset: Systematic evolution of the gap width, which closes at $\delta = 0.5$ for the dielectric system (blue circles), but only saturates to a non-zero value for the plasmonic system (red $\blacksquare$'s) even at highest disorder.}
\end{figure}

\begin{figure}[h]
	\includegraphics[width=8.5cm]{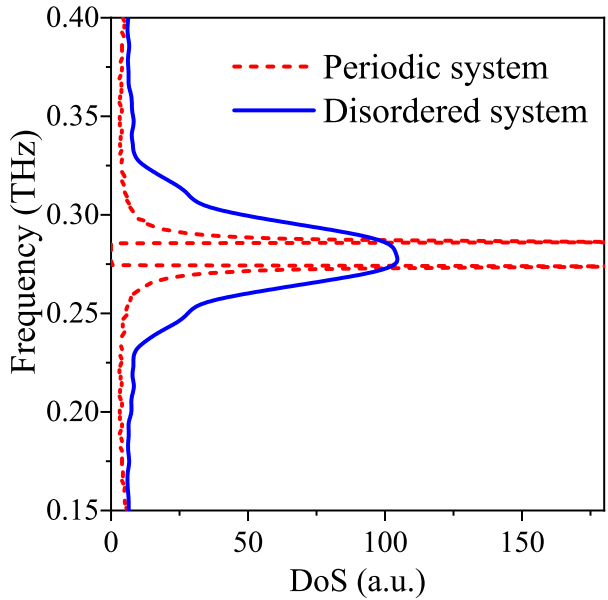}
	\caption{\label{fig:epsart9} Density of states of the hybrid plasmon system with strong disorder (solid blue line) in both, the cavity position and its  resonance. For comparison is shown the periodic limit (dashed red line). The DoS profile shows a maximum at the hybridisation frequency.}
\end{figure}

The behavior of the bands suggest an immediate influence on the density of states. For this analysis, we choose the first band in Fig. 1(a)). We choose the dielectric and plasmonic systems such that the plasmonic hybridization gap (Fig 8(a), red line) matches dielectric bandgap (blue line). The corresponding DoS for the dielectric system are depicted in Fig 8(b). In the periodic limit, Van Hove singularities occur at the bandedges (dotted lines). As the disorder increases, the modes migrate into the gap resulting in the Lifshitz tail (dashed blue line) and realize a finite non-zero DoS at the center of the gap. At strong disorder, the DoS has a flat profile. On the contrary, in the hybrid plasmonic system (Fig~8(c)), no flattening is seen at any disorder. The Van Hove singularities in the periodic limit are smoothened, but no such signature of a Lifshitz tail can be identified, and the DoS at the middle of the gap remains zero (dashed red line) at any disorder. The inset in Fig~8(c) shows the systematic evolution of the gap width with disorder. In the case of hybrid plasmons (red squares), the width of the gap never reaches zero, and in fact at strong disorder, the gap width asymptotically saturates to a finite value. In comparison, in the dielectric system (blue circles), the gap closes at $\delta = 0.5$.

Finally, we briefly address the situation wherein the cavities are not all alike, i.e., the resonant frequencies of the cavities themselves are random. This case is theoretically treated by randomising the $E_{cav}$ in Eq 2. Here, we only show the result at strong disorder. Specifically, at each diagonal site, an energy value was assigned by picking a uniform random number in the interval $[E_{cav} - 0.95\Delta E /2,E_{cav} + 0.95\Delta E /2]$. Here, the $\Delta E$ is the width of the hybridization gap. This resulted in diagonal disorder in addition to the earlier off-diagonal disorder.  The result is seen in Fig. 9, where the gap in the DoS profile is seen to close. But, the DoS are clearly enhanced at the hybridization frequency. As can be envisaged, the closure of the gap in the DoS is a systematic process such that at weaker disorder, the gap is still open(not shown). This behavior differs from conventional Anderson localization. Interestingly, we note that exactly such a situation arises in low-dimensional condensed matter systems such as graphene in presence of adatoms \cite{Garcia14}.

\section{Conclusions and Discussions}

In conclusion, we have theoretically studied Anderson localization in a hybrid-plasmonic structure consisting of a coupled array of sub-wavelength resonators. The system works at terahertz wavelengths, at which the high conductivity in metals allows for low losses. Therefore, systematic studies of simultaneous Anderson localization and dissipation can be carried out. We have approached the study starting with the band formation through hybridisation of surface plasmons and cavity resonances. Upon introduction of disorder, Anderson localized modes are observed at the edge of the hybridisation gap. The finite-element algorithm enables us to identify the loss length independently from the localization length. The latter is measured using wavefunction profile, which shows a clear exponential decay originating from the disorder. The theoretically observed behavior is consistent with our earlier experimental observation in a terahertz coupled cavity system, wherein Anderson localization was observed. We have further studied the regime of strong disorder, wherein previously unknown behavior was observed. Hitherto, most common studies on Anderson localization have been carried out in loss-less dielectric systems, wherein bandgaps are formed through direct or evanescent coupling of resonators, and the bandgap frequencies are distant from the resonant frequency. In the plasmonic case, however, the bandgap is centered at the cavity resonant frequency, which forces an unusual situation wherein the deepest modes in the hybridisation gap cannot transgress the cavity resonant frequency. This also manifests an interesting behavior of the density of states, in that the DoS profile never flattens despite the strong disorder. The number of modes at the center of the hybridisation gap remains zero, disregarding the degree of disorder. Even in systems with spoof plasmons such as at even lower frequencies, we expect the gap modes to behave similarly. To our knowledge, such behavior under Anderson localizing conditions is yet unstudied. It originates from the nature of the dispersion, which is similar to the Kondo-like hybridisation in metal crystals with magnetic impurities. It is of interest to investigate whether condensed matter systems with magnetic impurities exhibit similar consequences.

A natural progression of these studies concerns the observation of localization in two dimensions. Two-dimensional disordered systems are interesting due to the existence of diffuse-localized transition determined by the sample size, which does not exist in one-dimensional systems. Exponential decays can be measurable when the individual modes are isolated in space and frequency. From the above studies, there is a clear motivation to realize smaller system sizes in order to access the component modes of the system, that undergo localization under disorder. However, a trade-off will have to be made regarding the system size in order to cross the transition between light diffusion and Anderson localization. Taking into account the effect of aperture shapes on the transmission bands\cite{nahata04}, the subwavelength hole dimensions should be designed so as to maintain a clear separation in frequency between various bands so that the modes close to the bandedge remain accessible. An added factor to consider would be the metallic losses. These can be intuitively assumed to be maximum in the diffusion regime, given that the plasmonic field will have maximum spatial extent therein. In the localized regime, as seen from the above computations, the losses will reduce with increased disorder as the spatial extent reduces.

\begin{acknowledgments}
SM acknowledges funding from the Swarnajayanti Fellowship Grant, Department of Science and Technology, Government of India. AN acknowledges financial support by the NSF MRSEC program at the University of Utah under grant $\#~$DMR 1121252.

\end{acknowledgments}

\end{document}